\documentclass[prb,twocolumn,showpacs,amsmath,amssymb,superscriptaddress]{revtex4-1}
\usepackage{graphicx}
\usepackage{color}
\usepackage{braket}
\usepackage{amsmath}
\usepackage{amsfonts}
\usepackage{MnSymbol}
\usepackage{bm}

\usepackage{amsmath,amssymb,mathrsfs,ulem}
\usepackage[colorlinks = true,linkcolor = blue,urlcolor  = blue,     citecolor = blue,anchorcolor = blue]{hyperref}

\usepackage{bbm}
\usepackage[dvipsnames]{xcolor}
\usepackage{hyperref}

\date{\today}                  % Activate to display a given date or no date

\begin{document}

\title{Topological charge quantization on localized imperfections in crystalline insulators and the nearsightedness principle of Kohn}

\author{Kiryl Piasotski}
\email[Email: ]{kiryl.piasotski@kit.edu}
\affiliation{Institut f\"ur Theorie der Kondensierten Materie, Karlsruher Institut f\"ur Technologie, 76131 Karlsruhe, Germany}
\affiliation{Institut f\"ur QuantenMaterialien und Technologien, Karlsruher Institut f\"ur Technologie, 76021 Karlsruhe, Germany}
\thanks{On the leave from \textit{Institut f\"ur Theorie der Statistischen Physik, RWTH Aachen, 52056 Aachen, Germany}}

\begin{abstract}
We study the quantization of the excess charge on $N$ localized (ultra-screened) impurities in $d$-dimensional crystalline insulating systems. Solving Dyson's equation, we demonstrate that such charges are topological, by expressing them as winding numbers of appropriate functionals of bulk position space Green’s functions. We discuss the ties of our topological invariant with the nearsightedness principle of W. Kohn, stating that the electronic charge density at fixed chemical potential depends on the external field only locally, meaning that localized perturbations by external fields may only result in localized charge redistributions. We arrive at the same conclusion by demonstrating that an adiabatic perturbation comprised of a variation of impurities’ positions and/or strengths may only result in the change in the occupancy of impurity-localized bound states sitting, energy-wise, close to the Fermi level. Finally, we conclude by discussing the relations of the nearsightedness principle with the topological invariants characterizing the boundary charge.

\end{abstract}

\maketitle

\section{Introduction}

With the discovery of the quantum Hall effect [\onlinecite{Klitzing_1980}] and its topological origins [\onlinecite{Thouless_1982}, \onlinecite{Hatsugai_1993}] the study of the topological structures in condensed matter systems became a style rather than a fashion. The possibly best-known contemporary example of that is the field of topological insulators, where the boundary-localized edge states, with topologically granted existence and robustness, are being studied (popular reviews are [\onlinecite{Hasan_2010}, \onlinecite{Qi_2011}]). Despite being a well-defined endeavor with its very own periodic table [\onlinecite{Kitaev_2009}], this discipline leaves a number of questions open. One of these regards the direct experimental accessibility of these topological surface states. In particular, aside from the subspace of such states, their basis is incomplete for a description of the physical system accommodating them -- a topological insulator, making it highly questionable whether an actual physical observable may be expanded into a basis of intra- and inter-surface state transition operators. This is, for example, not true of the excess charge density that these insulators accumulate at their boundaries as it, as such, also features the exponentially localized contributions of all of the occupied extended states. Despite that, it is clear that as the surface states also contribute to such an observable, the change in their occupancy has to have an observable effect.

In a series of recent works [\onlinecite{Pletyukhov_2020}, \onlinecite{Muller_2021}, \onlinecite{Miles_2021}, \onlinecite{Pias_2022}], the topological properties of the boundary-localized electronic excess charges (the boundary charges) in unidimensional crystals were examined. In particular, a pair of topological invariants characterizing the boundary charge upon two bulk energy spectrum-preserving transformations of crystal's potential, translations and local inversions, were devised. Specifically, it was demonstrated that upon local inversion (inversion of coordinates within the unit cell), the boundary charge maps to its negative, up to an integral topological quantum number known as the interface invariant. Likewise, upon the lattice translation by $x_{\varphi}$, the boundary charge was shown to grow linearly with the shift variable $x_{\varphi}$ (with the slope being the unit cell-averaged average charge density in the bulk $\bar\rho=\frac{\nu}{L}$, with $\nu$ -- filling factor and $L$-system's period), whilst performing discontinuous downward jumps by a unit of the electron charge, as quantified by another topological quantum number -- the boundary invariant. These topological invariants were shown to be generated by the spectral flow of the energies corresponding to the edge states inside the energy gap that hosts the chemical potential, in complete analogy with the integer quantum Hall effect [\onlinecite{Hatsugai_1993}]. As opposed to the edge states in topological insulators, the quantization of these invariants does not rely on the internal symmetries of the bulk Bloch's Hamiltonian (such as particle-hole or time-reversal symmetries) and is instead guaranteed by a number of fundamental physical principles, such as charge conservation, Pauli principle, and the nearsightedness principle of W. Kohn [\onlinecite{Kohn_1978}, \onlinecite{Kohn_1996}, \onlinecite{Prodan_2005}, \onlinecite{Prodan_2006}] (to be discussed further on). Moreover, these invariants are directly linked with the properties of an experimental observable, a privilege shared by both the quantum Hall effect and the topological defects (see Ref. [\onlinecite{Trebin_1982}] for a review), while not being entirely clear in the domain of the topological insulators. 

Further, in a different paper [\onlinecite{Pletyukhov_2020_prr}], rational quantization of boundary and interface charges was discussed. Particularly, with the aid of the aforementioned physical principles, a general framework for studying quantized charges in one dimension was laid down, allowing us to quantify all possible quantization patterns of the boundary charge in terms of the non-symmorphic symmetries of the crystal. The charges on the interfaces between pairs of insulators sharing their bulk properties were demonstrated to follow a lattice version of the Goldstone-Wilczek formula [\onlinecite{Goldstone_1981}], relating the interface charge to the sum of the boundary charges right and left to their septum, modulo an unknown integer generated by the local coupling between the two subsystems. 

The key feature of the method developed in Ref. [\onlinecite{Pletyukhov_2020_prr}] is this ``modulo an unknown integer" paradigm, arising from the nearsightedness principle of the electronic matter. As such, the nearsightedness principle tells us that (see Ref. [\onlinecite{Prodan_2005}, \onlinecite{Prodan_2006}]), in insulators, localized perturbations by external fields may result in localized charge redistributions only. To be more specific, the corrections beyond the characteristic length scale $\xi_{g}=\frac{v_{F}}{E_{g}}$ (where $v_{F}$ and $E_{g}$ are the Fermi velocity and the gap opening up at the Fermi level, see Ref. [\onlinecite{Weber_2021}] for example) are exponentially suppressed. An even further refinement of this statement would be that such perturbations may only remove or add an additional number of bound states whose wave functions are localized around the corresponding perturbations. One of the key purposes of the present paper is to substantiate this claim mathematically, which turns out to be possible in pretty general $d$-dimensional models.

To be more specific, this paper concerns the topological properties of the electronic excess charges accumulated around point-like defects in $d$-dimensional insulators. Although we purposefully specify the Hamiltonian of the crystal under consideration to make our exposition more transparent, the derivations presented in this manuscript are shown to be independent of its choice. What indeed matters is that the spectrum of the clean system consists of the energy bands occasionally separated by the energy gaps, that is, there exists at least one bulk energy gap into which we can put the chemical potential to promote the resulting statistical system into an insulator. 

Furthermore, neither we specify the internal structure of the impurity vertices, nor do we assume any particular arrangement of them, making our analysis applicable to a wide range of experimental setups. In particular, quite conventionally, we may assume that a number of randomly located point-like impurities exerting an ultra-screened electrostatic force on the system's electrons are scattered through the charge sampling region of a crystal under consideration. A slightly less familiar situation is inspired by the work of Nomura and Nagaosa [\onlinecite{Nomura_2010}] and may be formulated as follows. Assuming that a crystal is further magnetic, we know that, in an insulating regime, its ground state may accurately be described by a Heisenberg model that, by itself, features topological defects. A familiar example of such a defect would be a magnetic skyrmion or a hedgehog texture. Assuming that the total spin of atoms comprising our crystal is large, these textures may be seen as an arrangement of classical magnetic moments nailed down to the atomic positions. Their interaction with the electron's spin degree of freedom may then be written as a sum of the Zeeman-like terms, each weighted with the Dirac $\delta$-function centered at the position of the corresponding atom. 

Quite generically, we show that the total electronic excess charge accumulated around these defects is an integer-valued topological invariant, which we express as a contour integral winding number of an appropriate functional of bulk position space Green's functions. Further analysis of this topological quantum number reveals that upon an adiabatic modification of positions and/or vertex functions of the localized scattering centers, the value of the invariant may only be affected by the change in the occupancy of the imperfection-localized bound states in the process of the spectral flow of their eigenenergies inside the chemical potential-accommodating energy gap. This observation allows for an immediate interpretation in terms of the nearsightedness principle, as well as for a direct read-off of the central memo of Ref. [\onlinecite{Pletyukhov_2020_prr}]: ``localized perturbations in insulators result in localized charge redistribution, leading to an addition/removal of the corresponding perturbation-localized bound states to/from the occupied spectral region". We conclude our analysis by commenting on the relation between the nearsightedness principle and the topological invariants characterizing the boundary charge.

In what follows, we set the reduced Plank's constant $\hbar$ and the electron charge $e$ equal to unity $\hbar=e=1$.

\section{Adiabatic response of the excess charge to localized perturbations in an insulating state}
\subsection{A translationally invariant model}
In the following, we shall specifically refer to an electronic system governed by the following Hamiltonian 
\begin{align}
    H^{(0)}_{\bold{x}}=\frac{\bold{p}^{2}}{2m}+\frac{1}{2m}\sum_{j=1}^{d}\{\tilde{A}_{j}(\bold{x}), p_{j}\}+V(\bold{x}), \label{Hamiltonian}
\end{align}
with $V(\bold{x})$ and $\tilde{A}_{j}(\bold{x}),\ j=1,\ \dots,\ d$ being the lattice periodic $N_{c}\times N_{c}$ Hermitian matrices. More specifically, 
\begin{align}
    \begin{Bmatrix}V(\bold{x}) \\  \tilde{\bold{A}}(\bold{x})\end{Bmatrix}=\begin{Bmatrix}V(\bold{x}+\bold{R}_{\bold{m}}) \\  \tilde{\bold{A}}(\bold{x}+\bold{R}_{\bold{m}})\end{Bmatrix}, \quad \forall \bold{m}\in\mathbb{Z}^{d},
\end{align}
where $\bold{R}_{\bold{m}}=\sum_{j=1}^{d}m_{j}\bold{a}_{j}$ is a lattice vector characterized by a $d$-dimensional vector of integers $\bold{m}=\begin{pmatrix}m_{1}& \hdots & m_{d}\end{pmatrix}^{T}$, specifying its components in the basis of primitive vectors $\{\bold{a}_{j}\}_{j}$ spanning the unit cell of a Bravais lattice. Furthermore, $\bold{p}$ and $\bold{x}$ are vectorial momentum and position operators comprised of the individual components $p_{j}=-i\frac{\partial}{\partial x_{j}}$ and $x_{j}$.

This model naturally generalizes the one recently studied in Ref. [\onlinecite{Pias_2022}] in connection with the universal properties of one-dimensional boundary charge, to higher dimensions. We remark that other models of multi-dimensional periodic structures [\onlinecite{other_models}] are expected to share the same physics, as the effects we are about to describe are rather generic to an insulating state. 

Translationally invariant systems are characterized by their band structure, comprised of the individual energy bands dispersing as $\epsilon_{\alpha, \bold{k}},\ \alpha=1,\ 2,\ \dots$, as a function of the vectorial quasimomentum variable $\bold{k}$, confined to the first Brillouin zone of the reciprocal space. The eigenstates of the Hamiltonian to which $\epsilon_{\alpha, \bold{k}}$ are the corresponding eigenvalues are known as Bloch functions $\psi_{\alpha, \bold{k}}(\bold{x})$, and may be generically expressed as
\begin{align}
\psi_{\alpha, \bold{k}}(\bold{x})=e^{i\bold{k}\cdot\bold{x}}u_{\alpha, \bold{k}}(\bold{x}),
\end{align}
where $u_{\alpha, \bold{k}}(\bold{x})$ in the $N_{c}$-component object and is lattice periodic in the same sense as vector and scalar potentials are $u_{\alpha, \bold{k}}(\bold{x})=u_{\alpha, \bold{k}}(\bold{x}+\bold{R}_{\bold{m}}), \ \forall \bold{m}\in\mathbb{Z}^{d}$. The completeness and identity resolution relations may be written as 
\begin{align}
\label{completeness}
    \frac{V_{\text{UC}}}{(2\pi)^{d}}\int_{\mathbb{R}^{d}}d^{(d)}\bold{x}\psi_{\alpha, \bold{k}}^{\dagger}(\bold{x})\psi_{\alpha', \bold{k}'}(\bold{x})&=\delta_{\alpha, \alpha'}\delta^{(d)}(\bold{k}-\bold{k}'),\\
\label{identity_resolution}
    \frac{V_{\text{UC}}}{(2\pi)^{d}}\sum_{\alpha=1}^{\infty}\int_{\text{BZ}}d^{(d)}\bold{k}\psi_{\alpha, \bold{k}}(\bold{x})\psi_{\alpha, \bold{k}}^{\dagger}(\bold{x}')&=1_{N_{c}}\delta^{(d)}(\bold{x}-\bold{x}'),
\end{align}
where $V_{\text{UC}}$ is the volume of the unit cell, defined via 
\begin{align}
    V_{\text{UC}}=\int_{\text{UC}}d^{(d)}\bold{x}=\det\begin{pmatrix}
        \bold{a}_{1} \Big{|}&\hdots & \Big{|}\bold{a}_{d}
    \end{pmatrix}.
\end{align}

When studying charge, it is more convenient to introduce the retarded single-particle Green's function, containing the information on both the eigenstates and the energy spectrum. In thermodynamic equilibrium, the Laplace image of the latter is defined as the resolvent of the single-particle Hamiltonian \eqref{Hamiltonian}
\begin{align}
    [z-H^{(0)}_{\bold{x}}]G^{(0)}(\bold{x}, \bold{x}')=1_{N_{c}}\delta^{(d)}(\bold{x}-\bold{x}'), \label{GF_def}
\end{align}
where $z$ is the complex energy variable, defined in terms of the physical frequency variable $\omega$ as $z=\omega+i\eta$, where $\eta\rightarrow 0^{+}$. Owing to the identity resolution relation \eqref{identity_resolution} we can establish the conventional Lehmann representation
\begin{align}
    G^{(0)}(\bold{x}, \bold{x}')=\frac{V_{\text{UC}}}{(2\pi)^{d}}\sum_{\alpha=1}^{\infty}\int_{\text{BZ}}d^{(d)}\bold{k}\frac{\psi_{\alpha, \bold{k}}(\bold{x})\psi_{\alpha, \bold{k}}^{\dagger}(\bold{x}')}{z-\epsilon_{\alpha, \bold{k}}}. \label{Lehmann}
\end{align}
Further, using the completeness of the basis \eqref{completeness}, in Appendix \ref{square_appendix}, we establish the following important fusion rule for the bare propagators
\begin{align}
   \int_{\mathbb{R}^{d}}d^{(d)}\bold{x}'G^{(0)}(\bold{x}, \bold{x}')G^{(0)}(\bold{x}', \bold{x}'')=-\frac{\partial}{\partial\omega}G^{(0)}(\bold{x}, \bold{x}'').
   \label{square_identity}
\end{align}
As it is shown in Appendix \ref{square_appendix}, this relation holds pretty generally, without any reference to the Hamiltonian \eqref{Hamiltonian}.

\subsection{Localized perturbations and Dyson's equation}
Now we perturb the translationally invariant (on the scale of the unit cell) system by a finite number of point-like impurities 
\begin{align}
\label{perturbation}
    \tilde{V}(\bold{x})=\sum_{n=1}^{N}\tilde{V}^{(n)}_{0}\delta^{(d)}(\bold{x}-\bold{x}_{n}),
\end{align}
where $\tilde{V}^{(n)}_{0}$ are $N_{c}\times N_{c}$ matrices describing the action of the $n^{\text{th}}$ impurity on the channel space. This action is further assumed to be local as prescribed by Dirac delta-function $\delta^{(d)}(\bold{x}-\bold{x}_{n})$ centered at the impurity position $\bold{x}_{n}$. 

Let us remark that the problem of a Dirac delta-function potential is well-known to be ill-defined in spatial dimensions higher than $d=1$. In our analysis, this is manifested in the ill-definiteness of the bulk position space Green's function at equal spatial arguments $G^{(0)}(\bold{x}, \bold{x})$ due to the divergence of the defining integrals \eqref{Lehmann} in the ultraviolet. Such a divergence is not physical and has to be circumvented by an appropriate regularization scheme. In particular, in the metallic case $\tilde{\bold{A}}(\bold{x})=0,\ V(\bold{x})=0$, in $d>1$ the problem of the delta-potential has been extensively studied in both physical [\onlinecite{Park_1995}, \onlinecite{Atkinson_1975}, \onlinecite{Jackiw_1991}] and mathematical [\onlinecite{Friedman_1972}] literature and several meaningful regularization techniques were proposed and shown to produce physically sensible results. Since the presence of the energy gaps is of no importance in the deep ultraviolet regime, the same methods may be applied in our case.

The Dyson's equation for the full Green's function of the system is given by
\begin{align}
\nonumber
    G(\bold{x}, \bold{x}')=&G^{(0)}(\bold{x}, \bold{x}')\\
    &+\sum_{n=1}^{N}G^{(0)}(\bold{x}, \bold{x}_{n})\tilde{V}^{(n)}_{0}G(\bold{x}_{n}, \bold{x}').
\end{align}
First we want to consistently solve for the functions $G(\bold{x}_{n}, \bold{x}'),\ n=1,\ \dots,\ N$. This problem is brought to the solution of the following matrix equation 
\begin{align}
\mathcal{M}(z)\mathcal{D}(\bold{x}')=\mathcal{D}^{(0)}(\bold{x}'),
\end{align}
where $\mathcal{M}(z)$ is the $N_{c}\cdot N\times N_{c}\cdot N$ block matrix defined by 
\begin{align}
\mathcal{M}(z)=&1_{N_{c}\cdot N}-\mathcal{G}^{(0)}(z)\tilde{\mathcal{V}}_{0},\\
(\mathcal{G}^{(0)}(z))_{n, n'}=&G^{(0)}(\bold{x}_{n}, \bold{x}_{n'}), \ (\tilde{\mathcal{V}}_{0})_{n, n'}=\delta_{n, n'}\tilde{V}_{0}^{(n)}. \label{important_matrices}
\end{align}
Likewise, $\mathcal{D}(\bold{x}')$ and $\mathcal{D}^{(0)}(\bold{x}')$ are the $N_{c}\cdot N\times N_{c}$ matrices comprised of the full $G(\bold{x}_{n}, \bold{x}')$ and bare $G^{(0)}(\bold{x}_{n}, \bold{x}')$ propagators, respectively. With these notations we obtain 
\begin{align}
\nonumber
    G(\bold{x}, \bold{x}')=&G^{(0)}(\bold{x}, \bold{x}')+\mathcal{D}^{(0)\dagger}(\bold{x})\tilde{\mathcal{V}}_{0}\mathcal{D}(\bold{x}')\\
    =&G^{(0)}(\bold{x}, \bold{x}')+\mathcal{D}^{(0)\dagger}(\bold{x})\tilde{\mathcal{V}}_{0}\mathcal{M}^{-1}(z)\mathcal{D}^{(0)}(\bold{x}'), \label{Dyson_matrix}
\end{align}
where in our definition the Hermitian conjugate does not affect the $z$-variable, i.e.
\begin{align}
    (G^{(0)}(\bold{x}, \bold{x}'))^{\dagger}=G^{(0)}(\bold{x}', \bold{x}).
\end{align}

\subsection{Measuring the excess charge}

We define the excess charge density operator in the following manner:
\begin{align}
    \delta\widehat{\rho}(\bold{x})=\widehat{\rho}(\bold{x})-\bar\rho, \label{excess_density}
\end{align}
where
\begin{align}
    \widehat{\rho}(\bold{x})=\widehat{\psi}^{\dagger}(\bold{x})\widehat{\psi}(\bold{x}),
\end{align}
is the density operator, expressed in terms of the $N_{c}$-component fermionic field operators $\widehat{\psi}(\bold{x})$ and $\widehat{\psi}^{\dagger}(\bold{x})$. The field operators $\widehat{\psi}(\bold{x})$ and $\widehat{\psi}^{\dagger}(\bold{x})$ are further assumed to destroy/create excitations of the full Hamiltonian including the effect of localized scattering centers in Eq. \eqref{perturbation}. The constant contribution $\bar\rho$ describes the unit cell-averaged average charge density in the bulk:
\begin{align}
    \bar\rho=&\frac{1}{V_{\text{UC}}}\int_{V_{\text{UC}}}d^{(d)}\bold{x}\rho^{(0)}(\bold{x}),\\
    \nonumber
    \rho^{(0)}(\bold{x})=&\braket{\widehat{\psi}^{(0)\dagger}(\bold{x})\widehat{\psi}^{(0)}(\bold{x})}\\
    =&-\frac{1}{\pi}\text{Im}\int^{\mu}_{-\infty}d\omega\text{tr}\left\{G^{(0)}(\bold{x}, \bold{x})\right\},
\end{align}
where the field operators $\widehat{\psi}^{(0)}(\bold{x})$ and $\widehat{\psi}^{(0)\dagger}(\bold{x})$ describe the excitations of the translationally invariant system, $\mu$ denotes the chemical potential, and $G^{(0)}(\bold{x},\ \bold{x}')$ is the bare Green's function defined by Eqs. \eqref{GF_def} and \eqref{Lehmann}.

We measure the excess charge with the help of the classical device, described by the envelope function $f(\bold{x})$ (see Refs. [\onlinecite{Pletyukhov_2020}, \onlinecite{Muller_2021}, \onlinecite{Miles_2021}, \onlinecite{Pias_2022}] and Ref. [\onlinecite{Kivelson_1982}] for similar definitions). To be more specific, we define the excess charge operator as
\begin{align}
    \delta\widehat{Q}=\int_{\mathbb{R}^{d}}d^{(d)}\bold{x}f(\bold{x})\delta\widehat{\rho}(\bold{x}).
\end{align}
It is sensible to define the function $f(\bold{x})$ relative to a certain point $\bold{x}_{p}$, to which the charge probe is applied, and further assume that the charge is sampled equivalently in all directions $f(\bold{x})=f(|\bold{x}-\bold{x}_{p}|)$. Additionally, we assume that all of the charge $f(|\bold{x}-\bold{x}_{p}|)\approx1$ is sampled in sufficiently large vicinity of the sampling point $\bold{x}_{p}$, while the envelope function smoothly decays to zero $f(|\bold{x}-\bold{x}_{p}|)\rightarrow 0$ far away from $\bold{x}_{p}$. For that matter, it is convenient to choose 
\begin{align}
f(|\bold{x}-\bold{x}_{p}|)=1-\Theta_{l_{p}}(|\bold{x}-\bold{x}_{p}|-L_{p}), \label{eq:envelope}
\end{align}
where $\Theta_{l_{p}}(|\bold{x}-\bold{x}_{p}|-L_{p})$ is some representation of the Heaviside function broadened by $l_{p}$. The length scales characteristic of the charge probe are assumed to satisfy 
\begin{align}
    L_{p}\gg l_{p}\gg \xi_{g},
\end{align}
where $\xi_{g}\simeq\frac{v_{F}}{E_{g}}$ is the charge localization length in an insulator (also it is the charge correlation length, defining the exponential decay length of the density-density correlation function, see Ref. [\onlinecite{Weber_2021}]), roughly defined as the ratio between the Fermi velocity $v_{F}$ and size of the energy gap at the Fermi level $E_{g}$. 

\subsection{Topological invariant characterizing the excess charge}
\label{top_invar_sec}

Let us assume that $N$ impurities, as characterized by the potential \eqref{perturbation}, are placed in a region of a crystal falling into the sampling district of the envelope function $|\bold{x}|\lesssim L_{p}$. We define the total excess charge as the zero temperature expectation value of the excess charge operator in the grandcanonical equilibrium density matrix, so that 
\begin{align}
    \delta{Q}=&\braket{\delta\widehat{Q}}=\int_{\mathbb{R}^{d}}d^{(d)}\bold{x}f(\bold{x})(\rho(\bold{x})-\bar\rho),\\
    \rho(\bold{x})=&-\frac{1}{\pi}\text{Im}\int_{-\infty}^{\mu}d\omega \text{tr}\left\{G(\bold{x}, \bold{x})\right\}.
\end{align}
With the help of the representation \eqref{Dyson_matrix}, we obtain 
\begin{align}
\delta{Q}=Q'+Q_{P},
\end{align}
where $Q'$ contains the Friedel charge as well as the charge due to the impurity-localized bound states
\begin{align}
Q'&=\int_{\mathbb{R}^{d}}d^{(d)}\bold{x}f(\bold{x})\rho'(\bold{x}),\\
\rho'(\bold{x})&=-\frac{1}{\pi}\text{Im}\int_{-\infty}^{\mu}d\omega \text{tr}\left\{\mathcal{D}^{(0)\dagger}(\bold{x})\tilde{\mathcal{V}}_{0}\mathcal{M}^{-1}(z)\mathcal{D}^{(0)}(\bold{x})\right\},
\end{align}
while $Q_{P}$ is the so-called polarization charge given by
\begin{align}
    Q_{P}=\int_{\mathbb{R}^{d}}d^{(d)}\bold{x}f(\bold{x})(\rho^{(0)}(\bold{x})-\bar\rho),
\end{align}
and, with the help of the properties of the envelope function, is shown to be zero $Q_{P}=0$ in Appendix \ref{Polarization_appendix}. It hence follows that 
\begin{align}
\nonumber
\delta{Q}=&Q'=-\frac{1}{\pi}\text{Im}\int_{\mathbb{R}^{d}}d^{(d)}\bold{x}f(\bold{x})\\
 &\times\int_{-\infty}^{\mu}d\omega\text{tr}\left\{\mathcal{D}^{(0)\dagger}(\bold{x})\tilde{\mathcal{V}}_{0}\mathcal{M}^{-1}(z)\mathcal{D}^{(0)}(\bold{x})\right\}.
\end{align}
Due to the branch cuts and poles of the $T$-matrix $T(\bold{x}, \bold{x}')=\sum_{n, n'}[\tilde{\mathcal{V}}_{0}\mathcal{M}^{-1}(z)]_{n, n'}\delta(\bold{x}-\bold{x}_{n})\delta(\bold{x}'-\bold{x}_{n'})$, the integrand of the outer integral is exponentially suppressed $\sim e^{-|\bold{x}|/\xi_{g}}$ at large $\bold{x}$, allowing us to set $f(\bold{x})=1$. 

Interchanging the order of the integrals, we consider 
\begin{align}
\nonumber
&\int_{\mathbb{R}^{d}}d^{(d)}\bold{x}\text{tr}\left\{\mathcal{D}^{(0)\dagger}(\bold{x})\tilde{\mathcal{V}}_{0}\mathcal{M}^{-1}(z)\mathcal{D}^{(0)}(\bold{x})\right\}\\
\nonumber
&=-\sum_{n, n'=1}^{N}\text{tr}\left\{[\mathcal{M}^{-1}(z)]_{n, n'}\frac{\partial}{\partial\omega}G^{(0)}(\bold{x}_{n'}, \bold{x}_{n})\tilde{V}_{0}^{(n)}\right\}\\
\nonumber
&=\sum_{n, n'=1}^{N}\text{tr}\left\{[\mathcal{M}^{-1}(z)]_{n, n'}\frac{\partial}{\partial\omega}[\mathcal{M}(z)]_{n', n}\right\}\\
\nonumber
&=\sum_{n=1}^{N}\text{tr}\left\{\left[\mathcal{M}^{-1}(z)\frac{\partial}{\partial\omega}\mathcal{M}(z)\right]_{n, n}\right\}\\
&=\frac{\partial}{\partial\omega}\text{tr}\left\{\log\mathcal{M}(z)\right\}=\frac{\partial}{\partial\omega}\log\det\left\{\mathcal{M}(z)\right\}, \label{intermediate_result}
\end{align}
where, in the last line, trace and determinant of the full $N_{c}\cdot N\times N_{c}\cdot N$ block matrix $\mathcal{M}(z)$ are understood. Using the result in Eq. \eqref{intermediate_result}, we arrive at the following compact formula for the total excess charge
\begin{align}
\delta{Q}=&-\frac{1}{\pi}\text{Im}\int_{-\infty}^{\mu}d\omega\frac{\partial}{\partial\omega}\log\det\left\{\mathcal{M}(z)\right\}. \label{inv_real_axis}
\end{align}

To see why the integral in Eq. \eqref{inv_real_axis} may take on integral values only, in Appendix \ref{cont_int_rep_appendix} we find an alternative contour integral representation
\begin{align}
\delta{Q}=&-\oint_{C}\frac{dz}{2\pi i}\frac{\partial}{\partial z}\log\det\left\{\mathcal{M}(z)\right\}, \label{top_inv_main}
\end{align}
where $C$ is an arbitrary non-self-intersecting curve that crosses the real axis at two points only, below the lowest eigenvalue of the full Hamiltonian and at the chemical potential $\mu$, and the direction of $C$ is assumed to be clockwise. 

In the representation \eqref{top_inv_main}, the excess charge $\delta{Q}$ is necessarily an integer as it is expressed as a contour integral winding number and the chemical potential is by definition inside one of the energy gaps (we focus on the insulating systems solely). In other words, the integral in Eq. \eqref{top_inv_main} measures the degree of the mapping $S^{1}\rightarrow S^{1}$ and is thus a member of the only non-trivial homotopy group of the unit circle $\pi_{1}(S^{1})=\mathbb{Z}$.

In particular, the integral in Eq. \eqref{top_inv_main}, is a sum of two distinct contributions: the contribution of the branch cuts corresponding to the extended or scattering states, and the contribution of poles corresponding to the imperfection-localized bound states. 

The bands in multidimensional ($d>1$) and/or multichannel ($N_{c}>1$) systems are typically composite, i.e. overlapping with one another along the frequency axis. For that matter, it is convenient to choose the branch cuts to connect the bottom of the lowest sub-band with the top of the highest one, within every patch of the energy bands surrounded by a pair of energy gaps.

The bound state poles, determined as a solution of $\det\left\{\mathcal{M}(z)\right\}\big{|}_{z\in\mathbb{R}}=0$, are located on the complement of the bare Hamiltonian's spectrum, i.e. inside the energy gaps and, in some cases (e.g. an attractive scalar impurity), below the bottom of the lowest energy band of the unperturbed Hamiltonian.

\section{Relation with the nearsightedness principle}
\begin{figure}
                \includegraphics[scale=0.20]{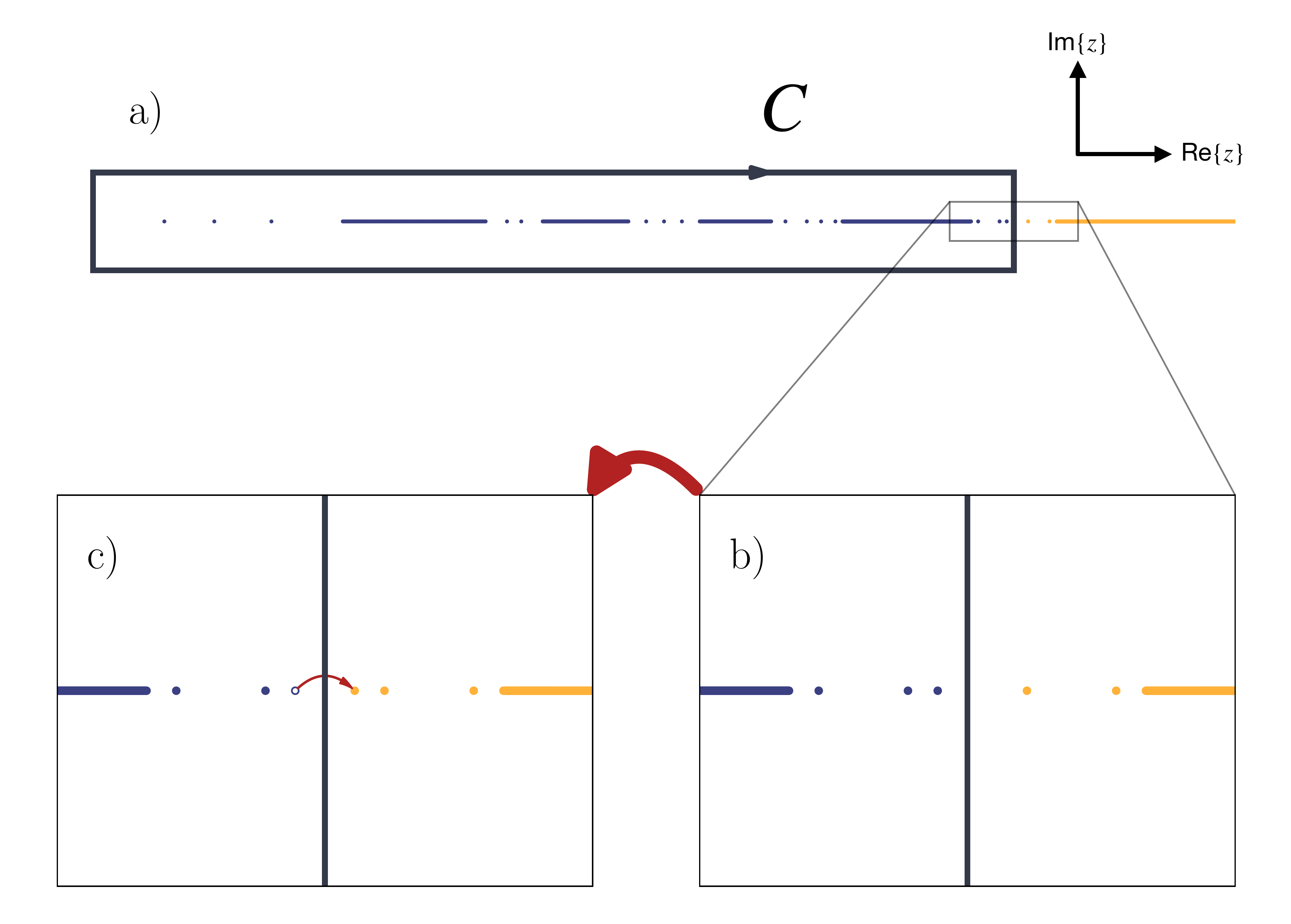}
                \caption{A schematic illustration of how the spectral flow of the energies of the imperfection-circumscribing bound states sitting inside the gap that accommodates the Fermi level affects the total excess charge. The spectrum of the system is visualized through the local spectral density as looked down on the complex frequency plane. The occupied part of the spectrum is demonstrated in blue, while the yellow color marks its complement (the states of the system that are unoccupied). Panel a) shows a rectangular contour $C$ encircling the occupied spectral region. Panels b) and c) show the zoomed-in vicinity of the chemical potential before and after the perturbation. As is demonstrated in panel c), the spectral flow results in the removal of a single bound state, carrying away a unity of the electron charge from the system (an inverse process is of course also possible). }
                \label{fig: spectral_flow_contour_scetch}
\end{figure}

\subsection{Discussion}
\label{Discussion}
Now we would like to discuss the topological invariant \eqref{top_inv_main} in greater detail. In what follows, we specify the contour $C$ as a rectangle of length $\mu-B$ in the real direction and width $2\eta$ in the imaginary one. Here $B$ is by definition an energy lying below the lowest eigenvalue of the total Hamiltonian $H_{\bold{x}}=H_{\bold{x}}^{(0)}+\tilde{V}(\bold{x})$ (i.e. $B\in(-\infty, \min\{\text{spec}\{H_{\bold{x}}\}\})$), and $\eta$ is not necessarily an infinitesimal positive but is rather a finite positive number (which is allowed as the integral is invariant under such contour deformations (see Appendix \ref{cont_int_rep_appendix})). Furthermore, we assume that the chemical potential is located above the $\nu^{\text{th}}$ bulk energy band.

Let us now consider making an adiabatic perturbation to the system that is comprised of the change in the positions $\{\bold{x}_{n}\}_{n}$ and/or vertex functions $\{\tilde{V}^{(n)}_{0}\}_{n}$ of the impurities. As the span of the extended states' energy bands is unaffected by such adiabatic perturbations, the branch cut contribution to the winding number remains invariant (up to the cases when the bound state merges with the band, as discussed below). This remark is essentially true as such deformations of the parameter space do not change the analytical structure of $G^{(0)}(\bold{x}_{n},\ \bold{x}_{n'})$, through the functionals of which alone our topological invariant is expressed. We hence conclude that such changes may only unleash themselves in the spectral flow of the bound state energies.     

As was anticipated in Section \ref{top_invar_sec}, the bound state energies are energy-wise located inside the energy gaps of the bulk system. This assertion also regards the energy gap below the bottom of the lowest band $\omega\in(-\infty,\ \min_{\bold{k}}\epsilon_{1, \bold{k}}]$, which can accommodate the bound states in the case of attractive impurities, for example. 

The energies of the bound states $\epsilon_{\text{bs}}$ inside the energy gaps $[\max_{\bold{k}}\epsilon_{\alpha, \bold{k}},\ \min_{\bold{k}}\epsilon_{\alpha+1, \bold{k}}]$ surrounded by a pair of bands $\alpha,\ \alpha+1,\ (\alpha=1,\ ...,\ \nu-1)$, are solely characterized by their location within the gap. The same holds true for the infinite gap below the bottom of the lowest bulk energy band, with $\epsilon_{\text{bs}}$ now being energy-wise located in $(-\infty,\ \min_{\bold{k}}\epsilon_{1, \bold{k}}]$. This implies that the spectral flow of these energies is constituted in the motion of $\epsilon_{\text{bs}}$ in between the top of $\epsilon_{\alpha, \bold{k}}$ and the bottom of $\epsilon_{\alpha+1, \bold{k}}$, or between the negative infinity and $\min_{\bold{k}}\epsilon_{1, \bold{k}}$ shall some states be also found in there. When merging with one of the energy bands (either $\epsilon_{\alpha, \bold{k}}$ or $\epsilon_{\alpha+1, \bold{k}}$, and $\epsilon_{1, \bold{k}}$ solely when considering the gap preceding the entire band structure), the value of the contour integral winding number \eqref{top_inv_main} relating to that band gets modified by unity \cite{footnote_2}. It follows that the motion of the bound state poles, inside such energy gaps below the one hosting the chemical potential, has absolutely no effect on the topological invariant \eqref{top_inv_main} (one may see this result as a form of charge conservation), as $B$, by definition, resides below the lowest pole (effectively meaning that none of the states are allowed to escape the occupied spectral region from below). 

The flow of the energies of the impurity-localized bound states residing inside the gap separating the conduction and the valence bands apart (the gap where the chemical potential is located), on the other hand, affects the winding number in Eq. \eqref{top_inv_main}. When a bound state crosses the chemical potential from above or below, the number of poles encompassed by the integration contour increases or decreases correspondingly. That means that the unit of the electron charge gets either pumped in or out of the system, modifying the topological invariant by $\pm1$. This discussion is summarized in Fig. \ref{fig: spectral_flow_contour_scetch}.

The elaboration above allows us to draw the following physical conclusion:
\begin{center}
    \textit{Localized adiabatic perturbations in insulators, may only result in the localized charge redistributions, owing to the change in the occupancy of the perturbation-localized bound states at the Fermi level.} 
\end{center}
This intuitive result is nothing but a direct consequence of the universal nearsightedness principle of W. Kohn [\onlinecite{Kohn_1996}, \onlinecite{Prodan_2005}, \onlinecite{Prodan_2006}] stating that, at fixed chemical potential, the electronic charge density depends on the external field (in our case being an assembly of localized scattering centers) only at nearby points. 

Another conclusion drawn by E. Prodan and W. Kohn in Ref. [\onlinecite{Prodan_2005}] (see also Ref. [\onlinecite{Prodan_2006}] for the fine details in $d=1$) is that the adiabatic perturbations to the external potential, no matter how strong, have a negligible effect on the local charge density beyond a certain characteristic length scale, which, in the insulating regime, is naturally provided by the charge correlation length $\xi_{g}$. From the viewpoint of our topological invariant \eqref{top_inv_main}, this means that in the case of well-separated impurities $|\bold{x}_{n}-\bold{x}_{n'}|/\xi_{g}\gg1$, the topological invariant is expected to approach a sum of the individual single-impurity invariants, as distant impurities are not supposed to be able to ``talk" with one another on such scales. Indeed, in an insulating state, it is well-known, that the two-point correlation functions $G^{(0)}(\bold{x}_{n},\ \bold{x}_{n'})$ decay exponentially at large distances $\sim e^{-|\bold{R}_{\bold{m}_{n}}-\bold{R}_{\bold{m}_{n'}}|/\xi_{g}}$ (where $\bold{m}_{n}$ labels the unit cell accommodating the $n^{\text{th}}$ scattering center), meaning that we can approximate 
\begin{align}
(\mathcal{G}^{(0)}(z))_{n, n'}\simeq&\delta_{n, n'}G^{(0)}(\bold{x}_{n}, \bold{x}_{n}),
\end{align}
implying that
\begin{align}
\mathcal{M}(z)\simeq&\bigoplus_{n=1}^{N}\left(1_{N_{c}}-G^{(0)}(\bold{x}_{n}, \bold{x}_{n})V_{0}^{(n)}\right),
\end{align}
and
\begin{align}
\delta{Q}\simeq&-\sum_{n=1}^{N}\oint_{C}\frac{dz}{2\pi i}\frac{\partial}{\partial z}\log\det\left\{1_{N_{c}}-G^{(0)}(\bold{x}_{n}, \bold{x}_{n})V_{0}^{(n)}\right\}. \label{approx}
\end{align}
This result may be seen as a form of the conventional Born approximation of the linear transport theory, whereby, to the lowest order in the impurity density, one considers impurities as independent. 

%Rigorous analysis of the post-Bornian (that are higher-order in the impurity density) corrections to Eq. \eqref{approx}, such as a weak-localization correction, is beyond the scope of the present manuscript and shall be performed elsewhere.  

\subsection{An illustration: A pair of magnetic impurities in an illuminated quantum wire}
\label{illustration_section}
\begin{figure*}
                \includegraphics[scale=0.33]{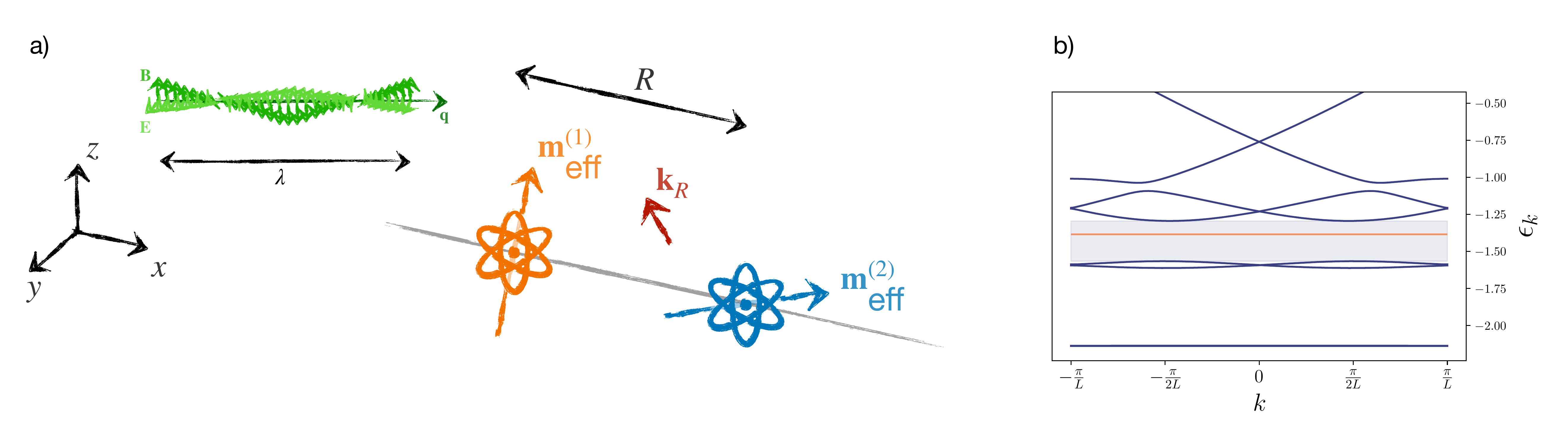}
                \caption{Panel a): A schematic illustration of a ballistic quantum wire featuring Rashba-style spin-orbit coupling (defined by a spin-orbit vector $\bold{k}_{R}$) and submersed into a spatially-periodic arrangement of electric $\bold{E}$ and magnetic $\bold{B}$ fields of wavelength $\lambda$. The two impurity atoms, separated by distance $R$ and carrying an effective magnetic moment of $\bold{m}_{\text{eff}}^{(j)},\ j=1,2$, are schematically shown by atomic symbols pierced with the magnetic moment-symbolizing arrows. Panel b): The bulk energy spectrum of the two-impurity problem. The energy bands are shown in dark blue, the chemical potential located inside the second spectral gap (above the fourth energy band) is depicted in orange, and the relevant spectral region is highlighted in light blue. }
                \label{fig: setup_sketch}
\end{figure*}
To illustrate some of the points highlighted in the above discussion, we here consider a simple model of a spin-orbit-interacting ballistic quantum wire, submersed into the background of the spatially oscillating electromagnetic field. The bulk Hamiltonian assumes the form of the Pauli Hamiltonian with an extra Rashba-like term:
\begin{align}
\nonumber
    H_{x}^{(0)}=&\frac{\left(p+\frac{e}{c}A_{x}(x)\right)^{2}}{2m}+\frac{\bold{k}_{R}\cdot{\bm{\sigma}}}{m}\left(p+\frac{e}{c}A_{x}(x)\right)\\
    &+\frac{\mu_{B}g_{e}}{2}{\bm{\sigma}}\cdot\bold{B}(x). \label{hamiltonian_example}
\end{align}
Above, $\bold{k}_{R}=(k_{R, x},\ k_{R, y},\ k_{R, z})$ is the Rashba spin-orbit vector, ${\bm{\sigma}}=(\sigma_{x},\ \sigma_{y},\ \sigma_{z})$ is the vector of the Pauli spin matrices, $\mu_{B}=\frac{e}{2mc}$ is the Bohr magneton, $c$ is the speed of light in vacuum, $g_{e}$ is the electron's Land\'e g-factor, and 
\begin{align}
    \bold{B}(x)=&\nabla\times\bold{A}(\bold{x})\Big{|}_{\bold{x}=x\hat{\bold{e}}_{x}},\quad A_{x}(x)=\hat{\bold{e}}_{x}\cdot\bold{A}(\bold{x})\Big{|}_{\bold{x}=x\hat{\bold{e}}_{x}},
\end{align}
with $\hat{\bold{e}}_{x}$ being the ort in the $x$-direction, and $\bold{A}(\bold{x})$ being the electromagnetic vector potential of the monochromatic plane-wave form
\begin{align}
    \bold{A}(\bold{x})=\bold{A}_{0}\cos(\bold{q}\cdot \bold{x}+\varphi),
\end{align}
in the Coulomb gauge 
\begin{align}
    \nabla\cdot\bold{A}(\bold{x})=0\iff \bold{q}\cdot\bold{A}_{0}=0.
\end{align}
The wave vector of the background electromagnetic field defines the fictitious lattice spacing 
\begin{align}
    L=\frac{2\pi}{\hat{\bold{e}}_{x}\cdot\bold{q}},
\end{align}
where we have excluded the uninteresting case of the orthogonally propagating wave $\hat{\bold{e}}_{x}\cdot\bold{q}=0$. 

We note that the Hamiltonian in Eq. \eqref{hamiltonian_example} falls into the class of systems defined by the Hamiltonian \eqref{Hamiltonian}, with $d=1$ and
\begin{align}
    V(x)=&\frac{\mu_{B}g_{e}}{2}{\bm{\sigma}}\cdot\bold{B}(x)+\frac{e^{2}A_{x}^{2}(x)}{2mc^{2}}+\frac{e\bold{k}_{R}\cdot{\bm{\sigma}}A_{x}(x)}{mc},\\
    \tilde{A}_{x}(x)=&\frac{e}{c}A_{x}(x)+\bold{k}_{R}\cdot{\bm{\sigma}}.
\end{align}

As this demonstration is assumed to be interpretative, it suffices to consider the case of a pair of impurities, which we assume to be separated by distance $R$:
\begin{align}
    \tilde{V}(x)=\tilde{V}^{(1)}_{0}\delta(x)+\tilde{V}^{(2)}_{0}\delta(x-R).
\end{align}
Note that we can place the first impurity at $x=0$ without loss of generality, as its other positions inside the wire may be achieved by appropriate tuning of the modulation's phase $\varphi$. Furthermore, we assume the impurities to exert both the electrostatic and the exchange ``force" on the wire's electrons, which we encode in the following form of the impurities' vertex functions
\begin{align}
    \tilde{V}^{(j)}_{0}=U_{j}\sigma_{0}+\frac{\mu_{B}g_{e}}{2}{\bm{\sigma}}\cdot\bold{B}_{\text{eff}}^{(j)},
\end{align}
where $\bold{B}_{\text{eff}}^{(j)}$ is the effective (also appropriately screened to have a short-ranged effect only) magnetic field, produced by the effective magnetic moment of the impurity atom $\bold{m}_{\text{eff}}^{(j)}=\frac{q_{j}g_{j}}{2M_{j}c}\bold{S}^{(j)}$, with $q_{j}$, $g_{j}$, and $M_{j}$ being the charge, g-factor, and mass of the $j^{\text{th}}$ impurity. Furthermore, $U_{j}$ denotes the strength of the electrostatic potential, defining the corresponding force exerted by the impurity on the electrons. Not going into much of the microscopic details, in the following, we treat $U_{j}$ and $\bold{B}_{\text{eff}}^{(j)}$ as some constant parameters. The resulting setup is schematically illustrated in panel a) of Fig. \ref{fig: setup_sketch}.

Further, to illustrate our point, we assume that the associated impurity parameters $\{R,\ \{U_{j}\}_{j},\ \{\bold{B}_{\text{eff}}^{(j)}\}_{j}\}$ evolve with a fictitious ``adiabatic time" $\tau\in[0, T]$, in such a manner that their temporal derivatives remain much smaller than the Fermi energy $\epsilon_{F}$ times their value, for all $\tau\in[0, T]$. 
\begin{figure*}
                \includegraphics[scale=0.145]{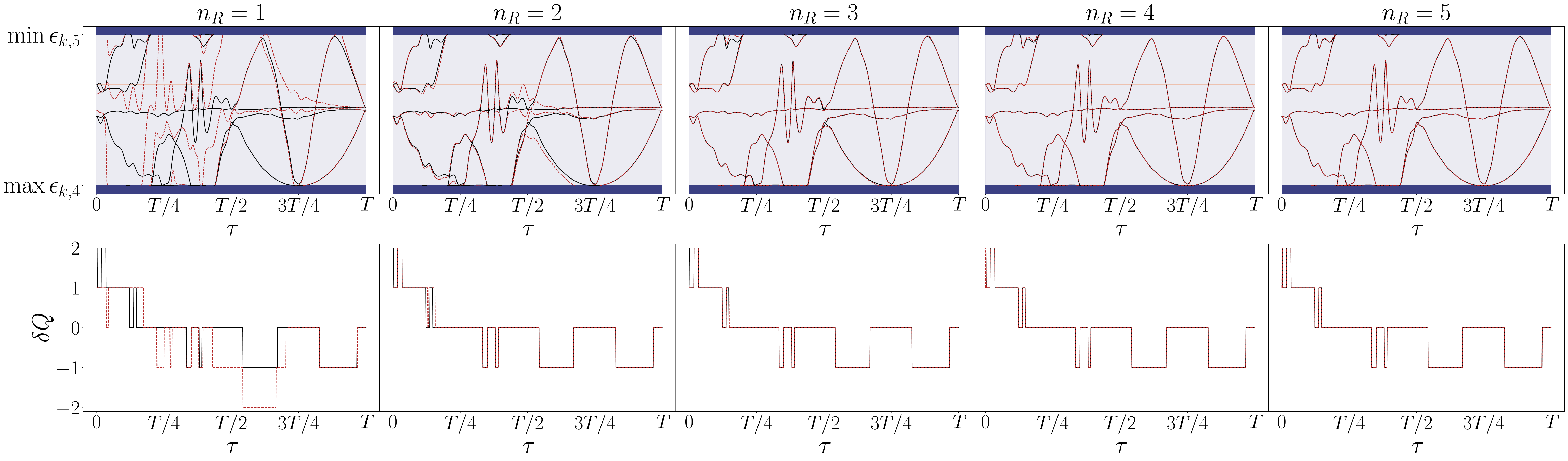}
                \caption{The figure demonstrates the adiabatic flows of both the bound state energy spectrum and the excess charge invariant in the toy model proposed in Section \ref{illustration_section}. Specifically, the spectral flow of the impurity-localized bound state energies is shown in the upper row, while the second row is dedicated to the invariant itself. As is explained in Appendix \ref{params_prots}, the position of the second impurity is parametrized as $R(\tau)=(n_{R}-1)L+\bar{R}(\tau)$, where $n_{R}$ denotes the number of the unit cell hosting the second imperfection, and $\bar{R}(\tau)\in[0, L]$ describes its location within the unit cell. The five distinct columns in the above figure correspond to five choices of $n_{R}=1,\ 2,\ 3,\ 4,\ 5$. In all of the panels, red dashed lines correspond to the actual solution, while black solid lines relate to the case of two independent impurities (see the approximate formula \eqref{approx}). As the separation between the impurities becomes of the order of the charge localization length $\xi_{g}=O(L)$ (see Appendix \ref{params_prots}), both adiabatic flows approach the limit of two independent impurities. }
                \label{fig: spectral_flow}
\end{figure*}

The particular form of the pumping protocol used to produce the numerical data and the concrete numerical values of the free model parameters are provided in Appendix \ref{params_prots}. The resulting bulk energy spectrum is demonstrated in panel b) of Fig. \ref{fig: setup_sketch}. 

The numerical data for the excess charge-characterizing topological invariant, as well as the spectral flow of bound state energies inside the chemical potential-accommodating spectral gap, is shown in Fig. \ref{fig: spectral_flow}. In particular, using the parametrization $R(\tau)=(n_{R}-1)L+\bar{R}(\tau)$, suggested in the Appendix \ref{params_prots}, we present the data for five different values of $n_{R}\in\{1,\dots, 5\}$, as shown in five different columns of the corresponding figure, with upper and lower rows corresponding to the spectral flow and the topological invariant, respectively. Solid black and dashed burgundy lines mark the cases of independent and ``interacting" impurities, correspondingly. By independent impurities, we here understand that the separation between them is effectively infinite, so that the off-diagonal blocks of the $\mathcal{M}(z)$ matrix (see Eq. \eqref{important_matrices} for the definition) may be completely ignored. This means that the bound state spectrum of the independent impurities is provided by the solution of $\det\left(1_{2}-G^{(0)}(0, 0)\tilde{V}^{(1)}_{0}\right)\det\left(1_{2}-G^{(0)}(R, R)\tilde{V}^{(2)}_{0}\right)\big{|}_{z\in\mathbb{R}}=0$, while the topological invariant is given by the approximation \eqref{approx}. By the ``interacting" impurities, on the other hand, we understand that the exact relations were used to produce the numerical data. The numerical technique for evaluation of bulk position space Green's functions, as well as the topological indices of the form \eqref{top_inv_main}, is outlined in Ref. [\onlinecite{Pias_2022}]. In our calculations, the values of the contour parameters were chosen as $\eta=1,\ B=-30$ (such a choice of $B$ is motivated by the presence of the bound states below the lowest band $\omega\in(-\infty,\ \min_{k}\epsilon_{k, 1}]$ in our model \eqref{hamiltonian_example}).

The central purpose of our demonstration is to show that upon the increase in the impurity's separation beyond the charge localization length $\xi_{g}=O(L)$ (see Appendix \ref{params_prots}), both the topological invariant and the bound state spectrum approach that of a pair of independent impurities. This effect is a direct consequence of the nearsightedness principle, telling us that a localized cause leads to a localized effect. Furthermore, as one may anticipate, the discontinuous jumps of the excess charge invariant occur precisely at the points where bound states enter/leave the occupied part of the energy spectrum, as is explained in Section \ref{Discussion}. Another interesting observation is the non-zero value of the topological invariant at the beginning of the adiabatic evolution in $\tau$, where the strengths of the electrostatic repulsion are the smallest $0<U_{j}\ll1$ (see Appendix \ref{params_prots}). This feature is a consequence of the presence of impurity-localized bound states below the bottom of the lowest energy band. Such an effect is well-known in the case of attractive scalar impurities, whereas here, it is generated by the non-Abelian structure of the model, and, to the best of our knowledge, was not reported previously in the literature.

\subsection{Topological invariants characterizing the boundary charge in unidimensional crystals} 

In this section, we would like to comment on the topological invariants characterizing boundary charges in unidimensional crystals, extensively discussed in Refs. [\onlinecite{Pletyukhov_2020}, \onlinecite{Muller_2021}, \onlinecite{Miles_2021}, \onlinecite{Pias_2022}]. In particular, let us consider a $d=1$ semi-infinite system described by the Hamiltonian \eqref{Hamiltonian}, with the boundary placed at $x=x_{b}$. An appropriate restriction of $x$ defines the respective right and left subsystems:
\begin{align}
    x&\in[x_{b}, \infty),\quad \text{right sub-system}, \label{restrict_1}\\
    x&\in(-\infty, x_{b}],\quad \text{left sub-system}. \label{restrict_2}
\end{align}
In our definition, the primitive unit cell is defined as the one starting at the boundary of the right semi-infinite system $\text{UC}=[x_{b},\ x_{b}+L]$, with $L$ being the lattice period. In this definition, the left half-system is always obtained from the right one by a local inversion operation, which acts by the inversion of local coordinates within each unit cell. 

Now we define the boundary charge operators corresponding to right and left semi-infinite systems as the envelope-weighted integrals of the expectation values of the appropriate excess charge density operators:
\begin{align}
    Q_{B}^{(R)}=&\int_{x_{b}}^{\infty}dxf(x)\braket{\delta\widehat{\rho}_{R}(x)},\\
    Q_{B}^{(L)}=&\int_{-\infty}^{x_{b}}dxf(x)\braket{\delta\widehat{\rho}_{L}(x)},
\end{align}
where, in analogy with Eq. \eqref{excess_density}, $\delta\widehat{\rho}_{S}(x)=\widehat{\rho}_{S}(x)-\bar\rho$, and $\widehat{\rho}_{S}(x)$ is the density operator referring to the system $S=R,\ L$. Furthermore, the envelope function $f(x)$ is chosen in accordance with Eq. \eqref{eq:envelope}, with $x_{p}=x_{b}$, and the range of $x$ being restricted according to Eqs. \eqref{restrict_1} and \eqref{restrict_2}. 

Let us now consider measuring the total excess charge $\delta {Q}$ accumulated around $x=x_{b}$ in a translationally invariant system $x\in(-\infty,\ \infty)$. By the polarization charge neutrality condition $Q_{P}=0$, demonstrated in Appendix \ref{Polarization_appendix}, the total excess charge also vanishes $\delta{Q}=0$. On the other hand, we may consider a translationally invariant system as a sum of right and left semi-infinite systems with a coupling corresponding to the bulk Hamiltonian switched in between them. This coupling manifests itself as a local perturbation and, by the nearsightedness principle of Kohn, is capable of affecting the total charge locally by at most introducing or removing a number of additional bound states, resulting in an integer contribution $Q_{I}$. In this connection, we conclude that $\delta{Q}=Q_{B}^{(R)}+Q_{B}^{(L)}-Q_{I}=0$, where $Q_{I}$ is known as the \textit{interface invariant}. One of the central results of Ref. [\onlinecite{Pias_2022}], was to demonstrate that
\begin{align}
    Q_{I}=&Q_{B}^{(R)}+Q_{B}^{(L)}=-\oint_{C}\frac{dz}{2\pi i}\frac{\partial}{\partial z}\log\det\left\{G^{(0)}(x_{b}, x_{b})\right\}.
\end{align}
That is, the interface invariant, characterizing the boundary charge upon local inversions, is a topological quantum number given by the winding of the determinant of bulk position space Green's function evaluated at the location of the boundary. 

Now let us proceed with the transformations of the boundary charge under translations. First, we consider the right boundary charge of the so-called reference system, starting at $x_{b}=0$:
\begin{align}
    Q_{B}^{(R)}(0)=\int_{0}^{\infty}dxf(x)(\rho(x)-\bar\rho),
\end{align}
and we would like to analyze the changes in this quantity upon the translation of the boundary by $x_{\varphi}\in[0,\ L]$. Instead of shifting the boundary, we consider adding the following potential $\hat{V}(x)=\hat{V}_{0}\Theta(x)\Theta(x_{\varphi}-x),\ \hat{V}_{0}\rightarrow\infty$. By the Pauli principle, the charge density becomes zero for $x\in[0, x_{\varphi}]$ as these states sit at infinite energy above the chemical potential $\mu$. From the definition of the boundary charge, we are left with the following contribution:
\begin{align}
\nonumber
    \delta Q_{B}^{(R)}(x_{\varphi})&=\int_{0}^{x_{\varphi}}dxf(x)(0-\bar\rho)\mod1\\
    &=-\bar\rho x_{\varphi}\mod1,
\end{align}
where $\mod1$ contribution again comes from the nearsightedness principle. This analysis allows us to conclude that:
\begin{align}
    Q_{B}^{(R)}(x_{\varphi})- Q_{B}^{(R)}(0)=\bar\rho x_{\varphi}+I(x_{\varphi}),
\end{align}
where $I(x_{\varphi})$ is known as the \textit{boundary invariant}. Another important result of Ref. [\onlinecite{Pias_2022}] was to show that 
\begin{align}
      & I (x_{\varphi})  = -\oint_{C}\frac{dz}{2\pi i}\frac{\partial}{\partial z}\ln \det  \mathcal{U} (x_{\varphi}),
\end{align}
where $\mathcal{U} (x_{\varphi})$ is defined via the path-ordered exponential 
\begin{align}
    \mathcal{U} (x_{\varphi})=&\text{P}\!\exp \left\{ \int_0^x d x' \mathcal{L} (x') \right\},\\
    \mathcal{L} (x)=&[G^{(0)} (x, x)]^{-1} G_2^{(0)} (x, x^+) - i A (x),
\end{align}
and $G_{2}^{(0)} (x, x')=\partial_{x'}G^{(0)} (x, x')$. In other words, the boundary invariant is also a topological quantum number expressed as a winding of the appropriate functional of bulk position space Green's functions.

In this way, we see that the quantization of the topological invariants characterizing the boundary charge in one-dimensional insulators is a direct consequence of the nearsightedness principle. As this intuitive physical principle holds beyond the single spatial dimension, one expects the excess charges accumulated on inhomogeneities of various spatial co-dimensions in $d$-dimensional crystals to possess similar topological characterization schemes. Indeed, linear scaling of the boundary charge, along with its discontinuous jumps by a unit of the electron charge at the bound state escape/entrance spectral points, was recently demonstrated in a two-dimensional system [\onlinecite{Hou_2022}].

\section{Conclusions and Outlook}
In this paper, the quantization of the excess charges on localized scattering centers in $d$-dimensional insulators was discussed. Our analysis reveals that an assembly of such imperfections accumulates an integral excess charge, given by a winding number expression. We find that an adiabatic perturbation (no matter how strong) comprised of either relocation of the impurities or a modification of their vertex functions (or both at the same time) results in the change of the total charge by an integer, determined by the saldo of the imperfection-localized bound states that entered or escaped the occupied spectral region, inside the chemical potential-hosting bulk spectral gap. The quantization of this topological invariant was shown to be a direct consequence of the nearsightedness principle of the electronic matter, limiting the range of the effect of a localized cause. Additionally, this local behavior of the electronic matter in the insulating state was shown to be responsible for the quantization of the topological invariants characterizing the unidimensional boundary charge studied in [\onlinecite{Pletyukhov_2020}, \onlinecite{Muller_2021}, \onlinecite{Miles_2021}, \onlinecite{Pias_2022}]. Furthermore, our study confirms the central paradigm of Ref. [\onlinecite{Pletyukhov_2020_prr}], namely that localized perturbations in insulators specifically lead to the change in occupancy of the corresponding perturbation-localized bound states, modifying the total charge, defined as the macroscopic average on the scales significantly exceeding both the unit cell size $L$ and the charge correlation length $\xi_{g}$, by at most an integer.

As is now obvious, the present paper is of conceptual value only as the evaluation of the suggested topological invariant \eqref{top_inv_main} for a specific multi-impurity ($N\gg1$) system poses a challenge on its own. In particular, this concerns questions regarding the regularization schemes for the higher-dimensional equal-argument Green's functions, as well as the basic questions regarding the numerical feasibility of the problem. Furthermore, it would be of future interest to study the expansion of the topological invariant in the interaction between the individual impurities, as generated by the off-diagonal blocks of the $\mathcal{M}(z)$ matrix, and analyze its ties with the conventional Born series for the impurity-dressed $T$-matrix. As it is suggested in the present study, in the insulating state, the impurity density $\rho_{I}$ has to be always contrasted with the inverse charge localization length $\xi_{g}$, in such a manner that the condition $1\gg \rho_{I}\xi_{g}^{d}$ implies the validity of the Born approximation, treating impurities as independent.

\section{Acknowledgments}
The author gratefully acknowledges the durable exchange of ideas with M. Pletyukhov and H. Schoeller. Further, the author generously thanks S. Miles and M. Pletyukhov for their valuable comments.

Most of the present work was done at the Institut f\"ur Theorie der Statistischen Physik of RWTH Aachen and was financially supported by the Deutsche Forschungsgemeinschaft via RTG 1995.

\begin{appendix}
\section{Contraction of two Green's functions}
\label{square_appendix}
Quite generically we may represent 
\begin{align}
    G=\sumint_{s}\frac{\ket{s}\bra{s}}{z-\epsilon_{s}},
\end{align}
where the meta-index $s$ labels the eigenstates $\ket{s}$ and eigenenergies $\epsilon_{s}$ of the Hamiltonian. Considering the product of the Green's function with itself 
\begin{align}
\nonumber
GG&=    \sumint_{s}\sumint_{s'}\frac{\ket{s}\bra{s'}}{(z-\epsilon_{s})(z-\epsilon_{s'})}\underbrace{\braket{s|s'}}_{\delta(s, s')}=\sumint_{s}\frac{\ket{s}\bra{s}}{(z-\epsilon_{s})^{2}}\\
&=-\frac{\partial}{\partial\omega}\sumint_{s}\frac{\ket{s}\bra{s}}{z-\epsilon_{s}}=-\frac{\partial}{\partial\omega}G.
\end{align}
Taking the position space matrix elements 
\begin{align}
\nonumber
\braket{\bold{x}|GG|\bold{x}''}&=-\frac{\partial}{\partial\omega}G(\bold{x}, \bold{x}''),
\end{align}
and inserting
\begin{align}
1=\int_{\mathbb{R}^{d}}d^{(d)}\bold{x}\ket{\bold{x}}\bra{\bold{x}},
\end{align}
we obtain the desired identity
\begin{align}
\int_{\mathbb{R}^{d}}d^{(d)}\bold{x}'G(\bold{x}, \bold{x}')G(\bold{x}', \bold{x}'')&=-\frac{\partial}{\partial\omega}G(\bold{x}, \bold{x}'').
\end{align}

\section{Polarization charge}
\label{Polarization_appendix}
We consider
\begin{align}
\nonumber
    &\int_{\mathbb{R}^{d}}d^{(d)}\bold{x}f(\bold{x})(\rho^{(0)}(\bold{x})-\bar\rho)\\
    &=\sum_{\bold{m}}\int_{\text{UC}}d^{(d)}\bar{\bold{x}}f(\bar{\bold{x}}+\bold{R}_{\bold{m}})(\rho^{(0)}(\bar{\bold{x}})-\bar\rho).
\end{align}
Above we parametrized the position space variable $\bold{x}$ as $\bold{x}=\bold{R}_{\bold{m}}+\bar{\bold{x}}$, for some vector of integers $\bold{m}$, and $\bar{\bold{x}}$ is the local coordinate within the unit cell $\bar{\bold{x}}\in\text{UC}$. Furthermore, we used the periodicity property of $\rho^{(0)}(\bold{x})$, implied by the periodicity of the equal-argument Green's function
\begin{align}
G^{(0)}(\bold{x},\ \bold{x})=G^{(0)}(\bold{x}+\bold{R}_{\bold{m}},\ \bold{x}+\bold{R}_{\bold{m}}),\quad \forall \bold{m}\in\mathbb{Z}^{d}.
\end{align}

The envelope function varies significantly only in the crossover region $|\bold{R}_{\bold{m}}|=O(L_{p})$, allowing us to approximate 
\begin{align}
f(\bar{\bold{x}}+\bold{R}_{\bold{m}})\approx f(\bold{R}_{\bold{m}})+\bar{\bold{x}}\cdot\nabla{f}(\bold{R}_{\bold{m}}),
\end{align}
leading to
\begin{align}
\nonumber
    &\int_{\mathbb{R}^{d}}d^{(d)}\bold{x}f(\bold{x})(\rho^{(0)}(\bold{x})-\bar\rho)\\
    &=\int_{\text{UC}}d^{(d)}\bar{\bold{x}}\sum_{\bold{m}}(\bar{\bold{x}}\cdot\nabla{f}(\bold{R}_{\bold{m}}))(\rho^{(0)}(\bar{\bold{x}})-\bar\rho).
\end{align}
Now approximating 
\begin{align}
\nonumber
&\sum_{\bold{m}}(\bar{\bold{x}}\cdot\nabla{f}(\bold{R}_{\bold{m}}))\approx \frac{1}{V_{\text{UC}}}\int_{\mathbb{R}^{d}}d^{(d)}\bold{y}(\bar{\bold{x}}\cdot\nabla_{\bold{y}}{f}(\bold{y}))\\
&=\frac{1}{V_{\text{UC}}}\int_{\mathbb{R}^{d}}d^{(d)}\bold{y}\nabla_{\bold{y}}\cdot(\bar{\bold{x}}{f}(\bold{y}))=0,
\end{align}
where in the last step we used Gauss' divergence theorem.

\section{Contour integral representation}
\label{cont_int_rep_appendix}
First, we rewrite Eq. \eqref{inv_real_axis} as
\begin{align}
\nonumber
\delta{Q}=&-\frac{1}{2\pi i}\int_{-\infty}^{\mu}d\omega\frac{\partial}{\partial\omega}\log\det\left\{\mathcal{M}(z)\right\}\\
&+\frac{1}{2\pi i}\int_{-\infty}^{\mu}d\omega\frac{\partial}{\partial\omega}\left(\log\det\left\{\mathcal{M}(z)\right\}\right)^{*}.
\end{align}
Now we remind ourselves that 
\begin{align}
\left(\log f(z)\right)^{*}=\log \left(f(z)\right)^{*}\equiv\log f^{*}(z).
\end{align}
Furthermore, one has 
\begin{align}
\left(\det\left\{\mathcal{M}(z)\right\}\right)^{*}=\det\left\{\mathcal{M}^{\dagger}(z^{*})\right\},
\end{align}
where, as before, the Hermitian conjugate does not affect the $z$-variable. Now we have 
\begin{align}
\nonumber
&\det\left\{\mathcal{M}^{\dagger}(z^{*})\right\}=\det\left\{\right(1-\mathcal{G}^{(0)}(z^{*})\tilde{\mathcal{V}}^{(0)}\left)^{\dagger}\right\}\\
\nonumber
&=\det\left\{1-\tilde{\mathcal{V}}^{(0)}\mathcal{G}^{(0)}(z^{*})\right\}\\
&=\det\left\{\mathcal{M}(z^{*})\right\},
\end{align}
where to get from the pre-last to the last lines we employed the Weinstein–Aronszajn identity. 

It hence follows that 
\begin{align}
\nonumber
\delta{Q}=&-\frac{1}{2\pi i}\int_{-\infty}^{\mu}d\omega\frac{\partial}{\partial\omega}\log\det\left\{\mathcal{M}(\omega+i\eta)\right\}\\
\nonumber
&-\frac{1}{2\pi i}\int_{\mu}^{-\infty}d\omega\frac{\partial}{\partial\omega}\log\det\left\{\mathcal{M}(\omega-i\eta)\right\}\\
=&-\oint_{C}\frac{dz}{2\pi i}\frac{\partial}{\partial z}\log\det\left\{\mathcal{M}(z)\right\}. \label{some_integral_third_appendix}
\end{align}
Above, $C$ is the counterclockwise rectangular contour defined as a union of four segments:
\begin{align}
\nonumber
    C=&[B+i\eta, \mu+i\eta)\cup[\mu+i\eta, \mu-i\eta)\cup[\mu-i\eta, B-i\eta)\\
    &\cup[B-i\eta, B+i\eta),\quad B\rightarrow-\infty,\quad \eta\rightarrow 0^{+}.
\end{align}

We note that the integral in \eqref{some_integral_third_appendix} remains unaffected under continuous contour deformations, so long as the analytic structure of the integrand within the patch of the complex plane enclosed by contour $C$ remains intact. In this connection, we may replace $C$ with an arbitrary non-self-intersecting curve crossing the real axis at two points only, at any energy below the lowest eigenvalue of the full Hamiltonian, and at the chemical potential. 

\section{Parameters and protocols}
\label{params_prots}
In the numerical example provided in Section \ref{illustration_section}, the parameters of the model were chosen according to
\begin{align}
    \bold{q}&=\frac{2\pi(\bold{e}_{x}+\kappa\bold{e}_{y})}{\lambda\sqrt{1+\kappa^{2}}},\quad \bold{A}_{0}=\frac{A_{0}(\kappa\bold{e}_{x}-\bold{e}_{y})}{\sqrt{1+\kappa^{2}}},\quad m=1,\\
    \kappa&=\frac{1+\sqrt{5}}{2}, \quad \lambda=4, \quad \frac{e}{c}A_{0}=1.17, \quad \bold{k}_{R}=\begin{pmatrix}
        0.32 \\ 1.39 \\ 1.24
    \end{pmatrix}.
\end{align}
Note that as we have set the electron's mass $m=1$ to unity (in addition to the electric charge $e=1$ and reduced Plank's constant $\hbar=1$), we work in Hartree's atomic units. 

In this way, the electromagnetic wave is propagating in the $x-y$ plane, with the corresponding magnetic field being 
\begin{align}
    \bold{B}(\bold{x})=\frac{2\pi A_{0}}{\lambda}\bold{e}_{z}\sin(\bold{q}\cdot\bold{x}+\varphi).
\end{align}
By definition, the corresponding lattice period is given by
\begin{align}
    L=\lambda\sqrt{1+\kappa^{2}}=2\sqrt{2(5+\sqrt{5})}.
\end{align}

To produce the data, we used the following pumping protocol for the impurities' separation
\begin{align}
    R(\tau)=(n_{R}-1)L+\bar{R}(\tau), \quad \bar{R}(\tau)=\frac{L}{T}\tau, \quad \frac{L}{T}\ll v_{F},
\end{align}
where $v_{F}$ is the Fermi velocity and $n_{R}$ is an integer specifying the number of the unit cell hosting the second impurity. For the impurities' vertex functions, we further make an assumption of the equivalent impurities: $U^{(1)}(\tau)=U^{(2)}(\tau)=:U(\tau)$ and $|\bold{B}_{\text{eff}}^{(1)}(\tau)|=|\bold{B}_{\text{eff}}^{(2)}(\tau)|=:B_{I}(\tau)$. The direction of the magnetic moments, on the other hand, is allowed to be different in two scattering centers and is parametrized in spherical polar coordinates
\begin{align}
    \frac{\bold{B}_{\text{eff}}^{(j)}(\tau)}{B_{I}(\tau)}=\begin{pmatrix}
        \cos\phi^{(j)}(\tau)\sin\theta^{(j)}(\tau) \\ \sin\phi^{(j)}(\tau)\sin\theta^{(j)}(\tau) \\ \cos\theta^{(j)}(\tau)
    \end{pmatrix}.
\end{align}

In the following, we assume that, as is the case with the location of the second impurity within the unit cell number $n_{R}$, the impurity strength also grows linearly with $\tau$
\begin{align}
    U(\tau)=U_{0}+\delta{U}\frac{\tau}{T},\quad \frac{\delta{U}}{T}\ll\epsilon_{F}^{2}.  
\end{align}
On the other hand, we assume the effective magnetic field of the impurity to oscillate as 
\begin{align}
    B_{I}(\tau)=B_{0}+\delta{B}\sin\left(\frac{6\pi \tau}{T}\right).
\end{align}
The direction of the spins is prescribed by 
\begin{align}
\phi^{(1)}(\tau)&=\phi^{(2)}(\tau)=2\pi\sin\left(\frac{8\pi\tau}{T}\right),\\
\theta^{(j)}(\tau)&=\frac{\pi}{2}\left(1+(-1)^{j}\frac{\tau}{T}\right).
\end{align}
The rest of the parameters are chosen as
\begin{align}
    U_{0}=0, \quad \delta{U}=10, \quad \frac{e}{c}B_{0}=3, \quad \frac{e}{c}\delta{B}=1.5.
\end{align}

Now let us estimate the charge localization length $\xi_{g}$ for the second bulk spectral gap, where the chemical potential $\mu$ is assumed to be placed. According to Ref. [\onlinecite{Pias_2022}], the Fermi velocity may roughly be estimated as $v_{F}\approx \frac{k_{F}}{m}\approx \frac{2\pi}{mL}\approx0.825816$. The energy gap at the Fermi level was numerically computed to be roughly $E_{g}\approx0.271394$, leading to the following estimate $\xi_{g}\approx 3=O(L)$.

\end{appendix}


\begin{thebibliography}{99}
\bibitem{footnote_2} In the case of one-dimensional ($d=1$) single-channel ($N_{c}=1$) systems, it is not possible for the bound state poles to coexist with the continuum (or extended) states. As a result, the effect of the bound state pole merging with the band is to increase the order of the branching pole at the band edge. In multichannel ($N_{c}>1$) and/or multidimensional ($d>1$) systems, on the other hand, the poles may, in principle, coexist with the band continuum. In either case, as a result, the band contribution to the winding number gets modified by unity.
\end{thebibliography}
\end{document}